# Identifying Participants in the Personal Genome Project by Name

Latanya Sweeney, Akua Abu, Julia Winn

Harvard College
Cambridge, Massachusetts
latanya@fas.harvard.edu, aabu@college.harvard.edu, jwinn@post.harvard.edu

*We linked names and contact information to publicly available profiles in the Personal Genome Project. These profiles contain medical and genomic information, including details about medications, procedures and diseases, and demographic information, such as date of birth, gender, and postal code. By linking demographics to public records such as voter lists, and mining for names hidden in attached documents, we correctly identified 84 to 97 percent of the profiles for which we provided names. Our ability to learn their names is based on their demographics, not their DNA, thereby revisiting an old vulnerability that could be easily thwarted with minimal loss of research value. So, we propose technical remedies for people to learn about their demographics to make better decisions.*

## INTRODUCTION

The freedom to decide with whom to share one's own medical and genomic information seems critical to protecting personal privacy in today's data-rich networked society. Individuals are often in the best position to make decisions about sharing extensive amounts of personal information for many worthy purposes like research. A person can weigh harms and benefits as relevant to her own life. In comparison, decisions by policy makers and committees do not usually allow fine-grained personal distinctions, but instead dictate sweeping actions that apply the same to everyone. But how good are the decisions individuals will make? A person may have far less expertise than vetted committee members or veteran policy makers. And potential harms and vulnerabilities may be hidden; if so, an individual may not be able to make good decisions.

For example, sharing information about sexual abuse, abortions, or depression medication may be liberating for one person yet harmful for another. Further, if the information is shared without the explicit appearance of name or address, a person may be more likely to share the information publicly because of a false belief she is anonymous.

It is important to help people make good data sharing decisions. If people share data widely and thousands of people get subsequently harmed doing so, policy makers may respond and take away the freedom to make personal data sharing decisions, thereby depriving society of individual choice. To make smarter decisions, people need an understanding of actual risks and ways technology can help.

## BACKGROUND

Launched in 2006, the Personal Genome Project (PGP) aims to sequence the genotypic and phenotypic information of 100,000 informed volunteers and display it publicly online in an extensive public database [1]. Information provided in the PGP includes DNA information, behavioral traits, medial conditions, physical characteristics, and environmental factors. A general argument for the disclosure of such information is its utility. The PGP founders believe this information will aid researchers in establishing correlations between certain traits and conducting research in personalized medicine. They also foresee its use as a tool for individuals to learn about their own genetic risk profiles for disease, uncover ancestral data, and examine biological characteristics [2]. According to the project's principal founder, Harvard geneticist George Church, the only real utility of this type of information is as data reflecting physical and genomic characteristics [3]. Along with Steven Pinker and Esther Dyson, he volunteered his information publicly as one of the first ten participants in the project. Currently, 2,593 individuals disclose their information publicly at the PGP website.

The PGP operates under a privacy protocol it terms "open consent"[4]. Individual volunteers freely choose to disclose as much personal data as they want, often including identifying demographic data, such as date of birth, gender, and postal code (ZIP). Online, the profiles appear in a "de-identified state," being void of the direct appearance of the participant's name or address. The result provides volunteers with seeming anonymity and a participant is assigned an identification number as the reference to his profile. Participants may upload information directly from external DNA sequencing services (e.g., from 23andMe), but these services often provide documents having additional personal information including the participant name.

PGP participants are required to sign a range of consent forms and pass an entrance exam. The consent form does not in any way guarantee participants a degree of privacy. To the contrary, the form explicitly states that participation may even reveal other non-disclosed information about the participant:

"If you have previously made available or intend to make available genetic or other medical or trait information in a confidential setting, for example in another research study, the data that you provide to the PGP may be used, on its own or in combination with your previously shared data, to identify you as a participant in otherwise private and/or confidential research. This means that any data or other information you may have shared pursuant to a promise of confidentiality or privacy may become public despite your intent that they be kept private and confidential. This could result in certain adverse effects for you, including ones not considered or anticipated by this consent form" [5]

Risks mentioned by the form include public disclosure and identification and the use of personal genomic information for non-medical purposes including cloning provided cell lines. It is emphasized that all risk lies with the individual.

Once a participant uploads information to his online profile, the PGP offers almost no means to amend or modify information. Participants basically display all the contents of the profile or none at all unless they know how to edit files directly. Some of these files use complicated and unusual formats (e.g., a continuity of care report that holds the participant's personal health record).

**Medical Privacy Regulation: HIPAA**

The Health Information Portability and Accountability Act (HIPAA) in the United States is the federal regulation that dictates sharing of medical information beyond the immediate care of the patient, prescribing to whom and how physicians, hospitals, and insurers may share a patient's medical information broadly. Patients are exempt. A patient may share his own information with whomever and wherever he chooses, and others have already posted detailed medical and genomic information – abortions, medications, and histories of abuse and mental illness – freely and publicly online at the PGP.

For medical data covered under HIPAA to be shared publicly beyond the control of the patient, dates would only contain the year, and the ZIP code would only have the first 2 digits if the population in the ZIP code is less than 20,000. No explicit identifiers, such as name, Social Security numbers, or addresses can appear.[1]

In sharp comparison, participants in the PGP often share their full date of birth (month, day and year) and the full 5 digit ZIP code, regardless of where they reside.

In 1997, Sweeney showed how demographics appearing in medical data that did not have the names of patients could be linked to registries of people (e.g., voter lists) to restore name and contact information to the medical data [6]. Her earliest example was identifying the medical information of William Weld, former governor of Massachusetts, using only his date of birth, gender, ZIP appearing in the medical data and a voter list [7]. Sweeney also used populations reported in the U.S. Census to predict that at most 87 percent of the U.S. population had unique combinations of date of birth, gender, and ZIP [6]. Recently, others have challenged whether there really is any vulnerability to being re-identified by date of birth, gender and ZIP, citing a lack of documented examples and being confused about whether Weld was re-identified because he was targeted or because his demographics were unique [8]; begging the question to be revisited. Can people be re-identified by date of birth, gender, and 5-digit ZIP?

**METHODS**

The database we used was 1130 public profiles copied from the PGP website [1] on September 1, 2011, which were all the public profiles available at the PGP at that time. About half of the profiles, 579 of 1130 or 51 percent, had date of birth, gender and 5-digit ZIP. Unless otherwise noted or is obvious from context, these 579 profiles comprise the base dataset for analysis ("Dataset").

**Experiment: Linking Demographics**

We conducted an experiment to determine how many of the profiles in Dataset we could re-identify by name using public records. Our sources for public records was a national sample of voter registrations ("Voter Data") and online access to a public records website ("Public Records"). The voter data was purchased from a third-party data broker and contained a sample of voter registrations for the 5-digit ZIP codes listed in Dataset. We estimate the sampling fraction in the Voter Data to be 72% based on a comparison of totals reported by state officials in one state (Massachusetts).

---

[1] 45 CFR 164.514(b)(2) (2002).

The approach involved matching demographics (date of birth, gender, ZIP) appearing in the PGP profiles of Dataset to those appearing in Voter Data or Public Records, and recording how many of the matches yielded just one name. Figure 1 shows a graphical depiction of our approach.

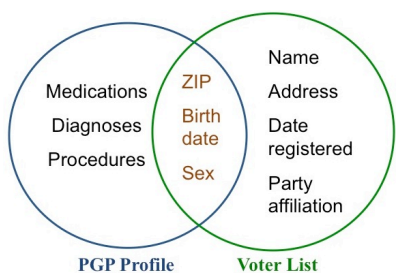

**Figure 1.** Linking PGP profile to a voter list using demographics to put names to the medical and genomic content appearing in the PGP profile.

**Experiment: Extracting Embedded Names**

We wrote programs in the Python and Java programming languages to download files from the publicly accessible PGP website; these were public profiles and other downloadable files associated with profiles. Some personal names appeared within downloaded documents. Our programs automatically extracted demographic information from the profile and names appearing in the filenames of compressed files that downloaded. From observation, we learned that many of the downloadable DNA files associated with PGP profiles were compressed, but once the file was uncompressed, the resulting file had a filename that included the name of the person as part of the filename. As a made-up example, the profile *hx0157A* may have a downloadable DNA file named *hx0157A_8659862.zip*, which when uncompressed gives the file having the name *genome_Elaine_Smith_Full_629562.txt*.

## RESULTS

Numerous tests were conducted on Dataset to demonstrate the ability to reliably put names to the profiles. Linking demographics in Dataset to those in the Voter Data yielded 130 (130 of 579 or 22 percent) unique matches. Our programs for finding embedded names located 103 (or 18 percent) names. Searches of the demographics from Dataset on a public records website yielded 156 (or 27 percent) unique matches. Combining these results gave a list of 241 (or 42 percent) unique names matching profiles. These were submitted to the PGP staff and scored: **84 percent correctly matched**, being as high as **97 percent if allowing considerations for possible nicknames** (e.g. Jim instead of James).

Table 1 describes the contributions of each strategy. Embedded names contributed 74 names not accounted for by the other strategies, linking to Voter Data contributed 44 distinct names and to Public Records, 65 names that would not otherwise have been known. Embedded names had 12 names in common with Voter Data and 17 with Public Records. Voter Data and Public Records had 74 names in common.

Table 2 reports the correctness of each strategy. The embedded names provided the highest number wrong (19 or 18 percent), primarily due to uses of nicknames. Voter Data matches gave the highest number of correct matches, even greater than Public Records, primarily because of a temporal problem inherent in the experimental design. Information stored in the profiles of Dataset was entered between 2006 and 2011. The Voter Data were as of 2011 but the Public Records were as of 2013. *A critical problem is believed to be the mobility of people from one ZIP code to another during those years, which causes a temporal mismatch between the Public Records and Dataset.*

|                    | Name | Voter | Public | *Totals* |
|--------------------|------|-------|--------|----------|
| **Name**           | 80   | 12    | 17     | *109*    |
| **Voter Data**     | 12   | 45    | 74     | *131*    |
| **Public Records** | 17   | 74    | 65     | *156*    |
| *Totals*           | *109*| *131* | *156*  |          |

**Table 1.** Discrimination of strategies. Values report the number of names specific to the strategy (e.g., embedded names contributing 74 names not otherwise found) or in common across strategies (e.g., 17 names found in both embedded names and Public Records).

|                    | Wrong | Total | Correct% |
|--------------------|-------|-------|----------|
| **Name**           | 19    | 103   | 82%      |
| **Voter Data**     | 9     | 130   | 93%      |
| **Public Records** | 20    | 156   | 87%      |

**Table 2.** Correctness of different re-identification strategies. Errors in matching embedded names and other strategies are due primarily to uses of nicknames rather than real names.

## DISCUSSION

These experiments demonstrate how PGP profiles are vulnerable to re-identification. What's the potential harm? Many participants reveal more than DNA, including seemingly sensitive conditions (e.g., profiles hu342A08, hu6D1115, and hu56B3B6) –abortions, sexual abuse, illegal drug use, alcoholism, clinical depression and more.

Perhaps more alarming are potential economic harms a participant may face. Here is an example. Suppose a hypothetical participant named Bob has a predisposition to a gene-based disease related to his genetic profile online. He applies for life insurance. If Bob is aware of the predisposition and discloses the information, he may be denied coverage or asked to pay a much higher premium. If he does not disclose knowledge of the predisposition or if he is not aware of the predisposition, the insurance company may fail to pay the claim upon his death. While the Genetic Information Non-Discrimination Act of 2008 (GINA) protects against some forms of discrimination (e.g. medical insurance), it does not cover all forms (e.g. life insurance).

Given the earlier discussion, we might have predicted being able to match unique names to more than 42 percent of the profiles. There are several possible explanations. The first is the temporal mismatch in the data described earlier. The second is our working with a sample of the voter data, rather than the entire voter file for each 5-digit ZIP. The third may be the quality of the data. And finally, Sweeney's prediction of 87% of the population being unique is an upper bound.

In concluding, what can a participant do to protect himself or herself? He can change the values in the fields that made this approach successful: *date of birth* or *ZIP code*. By making these values less specific, it becomes harder to link his name to the profile. Also, she can remove her name from appearing explicitly in documents she uploads.

To achieve these precautions required intervention on our part. We provide two technical services to PGP participants.

Sweeney constructed a website for a person to determine how unique his demographics may be (and therefore how easy it is to identify him) from his ZIP code, date of birth, and gender.[2] Anyone can check his or her demographics, even if not a participant in the PGP.

As stated earlier, the PGP itself does not support editing of the date of birth field, though participants can modify ZIP codes directly. So, we built a CCR (Continuity of Care Record) editor[3] for a participant to change his date of birth to report only year of birth or remove it altogether in the CCR file he uploads to the PGP.

Using knowledge from this study and associated services, individuals can make better data sharing decisions, or at least be more informed of risks and society can learn that date of birth, gender and 5-digit ZIP codes can be uniquely identifying.[4]

---

[2] The identifiability server is at http://aboutmyinfo.org/
[3] Available at https://mydatacan.hmdc.harvard.edu/pgp/
[4] Specific advice for PGP participants is available at http://dataprivacylab.org/projects/pgp

Acknowledgments

The authors gratefully thank Sean Hooley for help validating and scoring findings, participants in the Personal Genome Project (PGP) for being brave participants, and George Church, Jason Bobe and Madeline Ball for hosting the PGP. Our access to the PGP was limited to publicly available information. Julia Winn and Akua Abu did original re-identifications as class projects in the CS105 Privacy and Technology course at Harvard, taught by Jim Waldo and Latanya Sweeney. A copy of all data used in this study is available at foreverdata.org and the Dataverse Network. This work has been supported in part by a National Institutes of Health Grant (1R01ES021726).